\documentclass[amsmath,amssymb,aps,prl,reprint, showpacs, superscriptaddress]{revtex4-1}
\usepackage{graphicx}
\usepackage{amsthm}
\usepackage{amsmath}
\usepackage{amssymb}
\usepackage{dsfont}
\usepackage{bm}

\usepackage{color}
\begin{document} 
\title[]{Maximum Efficiency of Low-Dissipation Heat Engines at Arbitrary Power}
\author{Viktor Holubec}
\email{viktor.holubec@mff.cuni.cz}
\affiliation{ 
 Charles University in Prague,  
 Faculty of Mathematics and Physics, 
 Department of Macromolecular Physics, 
 V Hole{\v s}ovi{\v c}k{\' a}ch 2, 
 CZ-180~00~Praha, Czech Republic 
}
\author{Artem Ryabov}
\email{rjabov.a@gmail.com}
\affiliation{ 
 Charles University in Prague,  
 Faculty of Mathematics and Physics, 
 Department of Macromolecular Physics, 
 V Hole{\v s}ovi{\v c}k{\' a}ch 2, 
 CZ-180~00~Praha, Czech Republic 
}
\date{\today} 
\begin{abstract} 
We investigate maximum efficiency at a given power for low-dissipation heat engines. Close to maximum power, the maximum gain in efficiency scales as a square root of relative loss in power and this scaling is universal for a broad class of systems. For the low-dissipation engines, we calculate the maximum gain in efficiency for an arbitrary fixed power. We show that the engines working close to maximum power can operate at considerably larger efficiency compared to the efficiency at maximum power. Furthermore, we introduce universal bounds on maximum efficiency at a given power for low-dissipation heat engines. These bounds represent direct generalization of the bounds on efficiency at maximum power obtained by Esposito et al. Phys. Rev. Lett. {\bf 105}, 150603 (2010). We derive the bounds analytically in the regime close to maximum power and for small power values. For the intermediate regime we present strong numerical evidence for the validity of the bounds.  
\end{abstract}

\pacs{05.20.-y, 05.70.Ln, 07.20.Pe} 

\maketitle  

\section{Introduction}

Since the dawn of heat engines people struggle to optimize their performance \cite{Muller2007}. One of the first theoretical results in the field was due to Carnot \cite{Carnot1978} and Clausius \cite{Clausius1856}: The maximum efficiency attainable by any heat engine operating between the temperatures $T_{\rm h}$ and $T_{\rm c}$, $T_{\rm h} > T_{\rm c}$, is given by the Carnot efficiency $\eta_{\rm C} = 1 - T_{\rm c}/T_{\rm h}$. In order to attain $\eta_{\rm C}$, the engine must work reversibly (infinitely slowly) and thus its output power is vanishingly small. Optimization of the power of \emph{irreversible} Carnot cycles working under finite-time conditions was pioneered by Yvon \cite{Yvon1955}, Novikov \cite{Novikov1958}, Chambadal \cite{Chambadal1957} and later by Curzon and Ahlborn \cite{Curzon1975}. Although the obtained result for the efficiency at maximum power (EMP), $\eta_{\rm CA} = 1 - \sqrt{T_{\rm c}/T_{\rm h}}$, is not universal, neither it represents a bound on the EMP \cite{Hoffmann1997, Berry2000, Salamon2001}, its close agreement with EMP for several model systems \cite{DeVos1985,Bejan1996,JimenezdeCisneros2007,Schmiedl2008,Izumida2008,Izumida2009,Allahverdyan2008,Tu2008,Esposito2009a,Rutten2009,Esposito2010, Zhou2010,Zhan-Chun2012, Dechant2016} ignited search for universalities in performance of heat engines. 

Up to the second order in $\eta_{\rm C}$ the EMP, $\eta^\star$, is controlled by the symmetries of the underlying dynamics \cite{Esposito2009,Izumida2014,Sheng2015,Cleuren2015}. Further universalities were obtained for the heat engines working in the low-dissipation regime \cite{Esposito2010b, Sekimoto1997, Bonanca2014, Schmiedl2008, Tomas2013, Muratore-Ginanneschi2015}, where the work dissipated during the isothermal branches of the Carnot cycle grows in inverse proportion to the duration of these branches. In this regime, a general expression for the EMP has been published \cite{Schmiedl2008} and, subsequently, Esposito et al. derived the bounds $\eta_{\rm C}/2 \le \eta^\star  \le \eta_{\rm C}/(2 - \eta_{\rm C})$ on the EMP \cite{Esposito2010b}. All these results were confirmed within the framework of irreversible thermodynamics \cite{Izumida2012, Izumida2013}. 

Recently, increased attention has been given to the optimization of heat engines which does not work at maximum power \cite{Bauer2016, Holubec2015, Dechant2016, Whitney2014, Whitney2015}. Such studies are important for engineering practice, where not only powerful, but also economical devices should be developed. Indeed, it was already highlighted \cite{Chen2001,DeVos1992,Chen1994} that actual thermal plants and heat engines should not work at the maximum power $P^\star$, where the corresponding efficiency $\eta^\star$ can be relatively small, but rather in a regime with slightly smaller power $P$ and considerably larger efficiency $\eta$.

In the present paper, we introduce universal bounds on maximum efficiency at a given power for low-dissipation heat engines (LDHEs)
\begin{equation}
\frac{\eta_{\rm C}}{2}\left(1+\sqrt{-\delta_P}\right) \le \eta  \le \eta_{\rm C}\frac{1 + \sqrt{-\delta_P}}{2 - (1-\sqrt{-\delta_P})\eta_{\rm C}}\,,
\label{eq:deta_bounds}
\end{equation}
where
\begin{equation}
\delta_P = \left(P - P^{\star}\right)/P^{\star}\,.
\label{eq:dP_deta}
\end{equation}
We derive these bounds analytically for small $\delta_P$ and for $\delta_P$ close to $1$. For the intermediate regime we present strong numerical evidence that the bounds are valid for any $\delta_P$. The inequalities (\ref{eq:deta_bounds}) represent direct generalization of the bounds on EMP obtained for $\delta_P=0$ by Esposito et al. \cite{Esposito2010b}. In the leading order in $\eta_C$, the left and the right bound coincide and the resulting maximum efficiency, $\eta = \eta_{\rm C}(1+\sqrt{-\delta_P})/2$, equals to that obtained using linear response theory in the strong coupling limit \cite{Ryabov2016}. The both bounds coincide also for vanishing power ($\delta_P\to-1$), when they equal to $\eta_{\rm C}$, thus verifying Carnot's results.

We also study the maximum relative gain in efficiency 
\begin{equation}
\delta_{\eta} = \left(\eta - \eta^{\star}\right)/\eta^{\star}
\label{eq:deta}
\end{equation}
with respect to EMP of LDHEs \cite{Tomas2013, Long2015, Long2014, Sheng2013,Holubec2015} for arbitrary fixed power and show that it scales in the leading order of the relative loss of power $\delta_P$ as
\begin{equation}
\delta_{\eta} \propto \sqrt{-\delta_P}\,.
\label{eq:deta_dP}
\end{equation}
The slope of the gain in efficiency $\delta_{\eta}$ diverges at $\delta_P = 0$ and hence LDHEs working close to maximum power operate at considerably larger efficiency than $\eta^\star$. We show that, both the diverging slope and the scaling (\ref{eq:deta_dP}) are direct consequences of the fact that the maximum power corresponds to $\delta_P = 0$ and that these findings are valid  for broad class of systems [see the text below Eq.~(\ref{eq:efficiency_small_a})].
Indeed, the scaling (\ref{eq:deta_dP}) was already obtained in recent studies on quantum thermoelectric devices \cite{Whitney2014,Whitney2015}, for a stochastic heat engine based on the underdamped particle diffusing in a parabolic potential \cite{Dechant2016} and also using linear response theory \cite{Ryabov2016}.

\section{Model}
We consider a non-equilibrium Carnot cycle composed of two isotherms and two adiabats working in the low-dissipation regime \cite{Esposito2010b,Zulkowski2015a,Schmiedl2008,Zulkowski2015,Martinez2014, 
Blickle2011,Holubec2014, Rana2014, Rana2015,Benjamin2008,Tu2014,Holubec2015}. During the hot (cold) isotherm the system is coupled to the reservoir at temperature $T_{\rm h}$ ($T_{\rm c}$). Let $t_{\rm h}$ ($t_{\rm c}$) denotes the duration of the hot (cold) isotherm. In the low-dissipation regime, it is assumed that the system relaxation time is short compared to $t_{\rm h}$ and $t_{\rm c}$. Then it is possible to assume that the entropy production per cycle 
equals
\begin{equation}
\Delta S_{\rm tot} = \frac{A_{\rm h}}{t_{\rm h}T_{\rm h}} + \frac{A_{\rm c}}{t_{\rm c}T_{\rm c}}\,,
\label{eq:Stot}
\end{equation}
where $A_{\rm h}, A_{\rm c}$ are positive parameters. This means that the engine reaches reversible operation when duration of the cycle becomes very large ($t_{\rm h, \rm c}\to\infty$). Another usual assumption, that we also adopt here, is that the duration of the adiabatic branches is short compared to $t_{\rm h} + t_{\rm c}$ and thus the cycle duration can be well approximated by $t_{\rm p} = t_{\rm h} + t_{\rm c}$.

The heat absorbed by the system during the hot isotherm, $Q_{\rm h}$, and the heat delivered to the cold reservoir during the cold isotherm, $Q_{\rm c}$, are given by   
\begin{eqnarray}
Q_{\rm h} &=& T_{\rm h} \Delta S - A_{\rm h}/t_{\rm h}\,,
\label{eq:Qh}\\
Q_{\rm c} &=& T_{\rm c} \Delta S + A_{\rm c}/t_{\rm c}\,,
\label{eq:Qc}
\end{eqnarray}
where $\Delta S$ denotes the change of the system entropy during the hot isotherm. The positive parameters $A_{\rm h}$ and $A_{\rm c}$ thus measure the degree of irreversibility of the individual isotherms. They are given by the details of the dynamics of the system and can be easily measured \cite{Martinez2014}.

We express $t_{\rm h}$ and $t_{\rm c}$ using the duration of the cycle, 
$t_{\rm p}$, and its redistribution among the two isotherms, $\alpha$, as $t_{\rm h} = \alpha t_{\rm p}$ and $t_{\rm c} = (1-\alpha) t_{\rm p}$. Then the engine output power and its efficiency can be written as \cite{Holubec2014,Holubec2015}
\begin{eqnarray}
P &=& \frac{Q_{\rm h}-Q_{\rm c}}{t_{\rm p}} = \frac{(T_{\rm h} - T_{\rm c})\Delta S}{t_{\rm p}} - \frac{(1-\alpha)A_{\rm h} +  \alpha A_{\rm c}}{t_{\rm p}^2\alpha(1-\alpha)}\,,
\label{eq:power_Wirr}\\
\eta &=& \frac{Q_{\rm h}-Q_{\rm c}}{Q_{\rm h}} = \frac{\eta_{\rm C}}{1+T_{\rm c}\Delta S_{\rm tot}/(Pt_{\rm p})}\,.
\label{eq:eta}
\end{eqnarray} 

In general, interchanging the reservoirs at the ends of the isothermal branches brings the system out of equilibrium. During the subsequent relaxation, an additional positive contribution to the entropy production (\ref{eq:Stot}) arises, which may not vanish in the limit $t_{\rm h, \rm c}\to\infty$. This unavoidably results in a decrease of the efficiency at a fixed power (\ref{eq:eta}). By considering cycles with a reversible limit, we assume this dissipation to be negligible. While this assumption is reasonable for large systems, it might require a delicate control of system dynamics in case of microscopic heat engines \cite{Esposito2010b, Sato2002,Zulkowski2015a,Schmiedl2008,Zulkowski2015,Martinez2014,Holubec2014,Holubec2015}.

\section{Efficiency at maximum power}
Maximizing the power (\ref{eq:power_Wirr}) as the function of $t_{\rm p}$ and $\alpha$ gives \cite{Schmiedl2008} (values at maximum power are denoted by $\star$)
\begin{eqnarray}
\alpha^{\star} &=& \frac{A_{\rm h} - \sqrt{A_{\rm h}A_{\rm c}}}{A_{\rm h} - A_{\rm c}}\,,
\label{eq:alpha_opt}\\
t_{\rm p}^{\star} &=& \frac{2}{T_{\rm h}\eta_{\rm C}\Delta S}(\sqrt{A_{\rm h}}+\sqrt{A_{\rm c}})^2\,,
\label{eq:tp_opt}\\
P^{\star} &=& \frac{1}{4}\left(\frac{T_{\rm h}\eta_{\rm C}\Delta S}{\sqrt{A_{\rm h}} + \sqrt{A_{\rm c}}}\right)^2\,,
\label{eq:P_opt}\\
\eta^{\star} &=& \frac{\eta_{\rm C}(1+\sqrt{A_{\rm c}/A_{\rm h}})}{2(1+\sqrt{A_{\rm c}/A_{\rm h}})  - \eta_{\rm C}}
\label{eq:eta_opt}\,.
\end{eqnarray}
Note that the EMP (\ref{eq:eta_opt}) does not depend on the individual parameters $A_{\rm h}$ and $A_{\rm c}$, but only on their ratio $A_{\rm h}/A_{\rm c}$. 

\section{Efficiency near maximum power}
The operational point of maximum power (\ref{eq:alpha_opt})--(\ref{eq:eta_opt}) can be used to define the coordinate transformation 
\begin{align}
\tau &= \frac{t_{\rm p}}{t_{\rm p}^{\star}} - 1\,,& \tau &\in [-1,\infty)
\label{eq:tau}\,,\\
a &= \frac{\alpha}{\alpha^{\star}}-1\,,& a &\in [-1,\frac{1}{\alpha^{\star}} - 1]
\label{eq:a}\,,
\end{align}
which decreases the number of parameters contained in the formulas (\ref{eq:power_Wirr})--(\ref{eq:eta}) for power and efficiency by 2 \cite{Holubec2015}
and thus makes the maximization of efficiency for a given power much easier. The point of maximum power corresponds in these coordinates to the origin, i.e., $\tau=a=0$. The parameter $\tau$ is larger than zero whenever $t_{\rm p} > t_{\rm p}^{\star}$ and similarly $a > 0$ if $\alpha > \alpha^{\star}$.

The relative loss of power (\ref{eq:dP_deta}) and the relative change in efficiency (\ref{eq:deta}) in these new coordinates read
\begin{align}
\delta_P &= \frac{a^2}{(1 + a) (a - A) (1 + \tau)^2} - \left(\frac{\tau}{1 + \tau}\right)^2
\,,\label{eq:reldP}\\
\delta_{\eta} &= - 1 + \frac{2(1+A)-\eta_{\rm C}}{a - A} \times\\
&\times \frac{a (2 a - A + 1) - A + 2 (1 + a) (a - A) \tau}{2(1 + \tau)(1 + a)(1 + A) - \eta_{\rm C}}\,,
\label{eq:reldeta}
\end{align}
where 
\begin{equation}
A = \sqrt{A_{\rm c}/A_{\rm h}}\,. 
\label{eq:A}
\end{equation}
Let us here stress that by using the symbol $\delta$ in the notation we do not mean that the deviations from the maximum power measured by the functions (\ref{eq:reldP}) and (\ref{eq:reldeta}) must be small.

The power exhibits maximum at $\tau=a=0$ and thus $\delta_P$ for small $\tau$ and $a$ varies very slowly. On the other hand, the efficiency can change much more rapidly and thus, for suitable parameters, the loss of power is much smaller than the gain in efficiency \cite{Chen2001,DeVos1992,Chen1994, Holubec2015, Whitney2014, Whitney2015, Dechant2016}. We will now find the formula which describes this gain. 

\section{Maximum gain in efficiency for a fixed loss of power}
\begin{figure}
	\centering
		\includegraphics[width=1.0\linewidth]{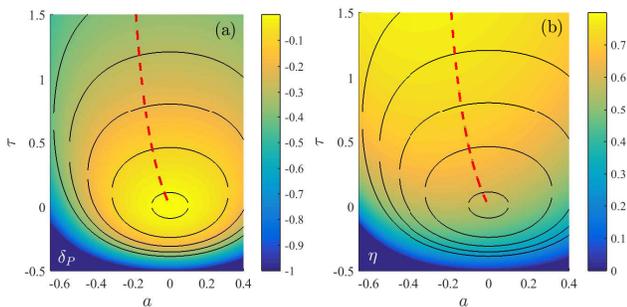}
	\caption{The relative loss of power (\ref{eq:reldP}) [panel (a)] and the efficiency $\eta = \eta^\star(\delta_\eta+1)$ [panel (b)] as functions of the parameters $a$ and $\tau$. In the both panels, the upper black lines were calculated from Eq.~(\ref{eq:tau_exact}) with the upper sign. Similarly, for calculation of the lower black lines we have used Eq.~(\ref{eq:tau_exact}) with the lower sign. These lines connects the points with the same value of $\delta_P$. The red dashed lines correspond to the maximum efficiency for a fixed power, which is the parameter of this curve. In the both panels we set $A = 1$, $\eta_{\rm C} = 0.875$.}
	\label{fig:dP_eta}
\end{figure}

For a fixed $\delta_P$, the parameters $a$ and $\tau$ are related due to Eq.~(\ref{eq:reldP}) as
\begin{equation}
\tau = \frac{-\delta_P}{1+\delta_P} \pm \frac{\sqrt{-a^2-[(1+a)A-a]\delta_P}}{\sqrt{(1+a)(A-a)}(1+\delta_P)}\,.
\label{eq:tau_exact}
\end{equation}
For five values of $\delta_P$ and for $A=1$, the curves defined by Eq.~(\ref{eq:tau_exact}) are depicted by black lines in Fig.~\ref{fig:dP_eta}. Upper (lower) lines correspond to the upper (lower) sign on the right-hand side of Eq.~(\ref{eq:tau_exact}). They mark the combinations of coordinates $a,\tau$ which yield the same value of power. The power is the larger the closer the curves are to the origin $a=\tau=0$. In this figure, we also show the relative loss of power $\delta_P$ [panel (a)] and the efficiency $\eta$ [panel (b)] as functions of the parameters $a$ and $\tau$. 

\begin{figure}
	\centering
		\includegraphics[width=1.0\linewidth]{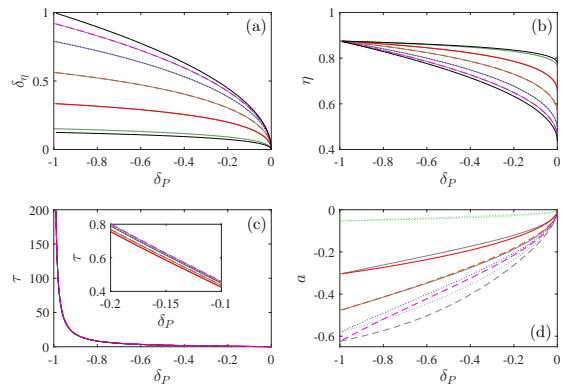}
	\caption{Panel (a): the maximum relative gain in efficiency $\delta_{\eta}$ (\ref{eq:deta}) as a function of the relative loss of power $\delta_{P}$ (\ref{eq:dP_deta}) for $\eta_{\rm C} = 0.875$ and five values of the parameter $A$: $A = \sqrt{0.001}$ (green dotted line), $A = \sqrt{0.1}$ (red solid line), $A = 1$ (black orange line), $A = \sqrt{10}$ (blue dotted line) and $A = \sqrt{100}$ (magenta dashed line). The dashed (full) black lines depict the lower (upper) bound on the maximum relative gain in efficiency (\ref{eq:deta_bounds2}). The corresponding efficiencies together with the bounds (\ref{eq:eta_bounds}) are shown in the panel (b). In panels (c) and (d) we
	show the corresponding optimal values of the parameters $\tau$ and $a$. The colored lines are calculated using exact numerical optimization of efficiency for a fixed power.
	The thin gray lines are calculated using analytical optimization based on the approximate formula Eq.~(\ref{eq:t_small_a}). Although the optimal values of the parameter $a$ calculated in this approximation sometimes differ from the correct values [panel (d)], the resulting optimal parameter $\tau$ [panel (c)] and, more importantly, the optimal 
	efficiency [panel (b)] and the optimal gain in efficiency [panel (a)] are predicted so well that the individual gray and colored curves overlap.}
	\label{fig:universal}
\end{figure}

\subsection{Exact numerical results}
\label{sec:Num_res}

Due to the algebraic complexity of Eqs.~(\ref{eq:reldP}) and (\ref{eq:reldeta}), the analytical derivation of the maximum $\delta_\eta$ for a given $\delta_P$ is in general intractable and we perform this calculation only numerically. Examples of the results of such optimization are demonstrated in Figs.~\ref{fig:dP_eta} and \ref{fig:universal}. In Fig.~\ref{fig:dP_eta} the dashed red line denotes the values of $a$ and $\tau$ corresponding to the maximum efficiency (and thus also $\delta_\eta$) for given values of $\delta_P$, which is the parameter of this curve. The dashed line intersects the upper solid black curves. Hence the optimal values of efficiency are obtained for the upper sign in Eq.~(\ref{eq:tau_exact}). In Fig.~\ref{fig:universal} we show the maximum gain in efficiency for a fixed power [panel (a)], the maximum efficiency for a fixed power [panel (b)] and the corresponding optimal~\footnote{In the following we will use the word `optimal' as a synonym for `corresponding to the maximum efficiency for a fixed power'.} values of the parameters $\tau$ [panel (c)] and $a$ [panel (d)] as functions of the relative loss of power $\delta_P$. The panels (a) and (b) in Fig.~\ref{fig:universal} demonstrate that the gain in efficiency when working close to maximum power ($\delta_P=0$) is indeed significant. The panels (c) and (d) in Fig.~\ref{fig:universal} and the red dashed line in Fig.~\ref{fig:dP_eta} reveal that optimal values of $\tau$ are always positive ($t_{\rm p}>t_{\rm p}^\star$) and the optimal values of $a$ are always negative ($\alpha < \alpha^\star$). This result is quite intuitive. 

For a fixed power, the efficiency (\ref{eq:eta}) increases if the average entropy production rate during the cycle, $\Delta S_{\rm tot}/t_{\rm p}$, decreases. For a fixed $\alpha$, the total entropy production per cycle $\Delta S_{\rm tot}$ (\ref{eq:Stot}) decreases with increasing $t_{\rm p}$ (and thus $\Delta S_{\rm tot}/t_{\rm p}$ decreases even faster). Physically, this is because slower processes are more reversible. On the other hand, for a fixed $t_{\rm p}$, $\Delta S_{\rm tot}$ is smaller for $\alpha < \alpha^\star$ than for $\alpha \ge \alpha^\star$. To see this, let us expand $\Delta S_{\rm tot}$ into a Taylor series around the point of maximum power $\alpha^\star$: $\Delta S_{\rm tot}  = \Delta S_{\rm tot}^\star +  (\Delta S_{\rm tot}')^\star (\alpha-\alpha^\star) + O[(\alpha-\alpha^\star)^2]$, where $\Delta S_{\rm tot}^\star$ is the total entropy production at maximum power and $(\Delta S_{\rm tot}')^\star = (1+A)^2 A_{\rm h}(T_{\rm h} - T_{\rm c})/(T_{\rm c}T_{\rm h}t_{\rm p}) > 0$. We thus have $\Delta S_{\rm tot} - \Delta S_{\rm tot}^\star < 0$ whenever $\alpha<\alpha^\star$. Although this prove is valid only up to the linear order in $\alpha-\alpha^\star$, the result holds generally. In order to get further physical intuition it is helpful to consider the symmetric situation $A=1$. In such case for smaller $\alpha$ (larger $1-\alpha$) more amount of work is dissipated in the bath with the large temperature $T_{\rm h}$, where the same amount of dissipated work creates less entropy than it would generate in the cold bath [entropy produced in a bath equals to (energy delivered to the bath)/(bath temperature)]. For a fixed power, $\alpha$ and $t_{\rm p}$ ($a$ and $\tau$) can not change independently and thus a compromise between an increased $\tau$ and a decreased $\alpha$ which verify Eq.~(\ref{eq:tau_exact}) is chosen. In this compromise, depicted in Fig.~\ref{fig:dP_eta} by the dashed red line, increasing $\tau$ makes the cycle more reversible and decreasing $\alpha$ causes that more energy is dissipated in the hot bath, which generates less entropy.

\subsection{Approximate analytical results}

Although the full analytical optimization of efficiency for a fixed power is in general beyond our reach, there are two limiting regimes when the analytical calculation is possible. The resulting simple analytical formulas (\ref{eq:deta_largedP}), (\ref{eq:deta_small_a}) and (\ref{eq:efficiency_small_a}) yield the bounds (\ref{eq:deta_bounds2}) and (\ref{eq:eta_bounds}) on maximum $\delta_\eta$ and $\eta$ for a fixed power. Comparison with exact numerics reveals that these bounds are valid also outside the two limiting regimes (see Fig.~\ref{fig:universal} and explanations below).

First, for $\delta_P\to -1$ ($P\to 0$), Eq.~(\ref{eq:tau_exact}) yields $\tau\to\infty$ ($t_{\rm p}\to\infty$). Then, we get from Eq.~(\ref{eq:reldeta}) that 
\begin{equation}
\delta_\eta = \frac{\eta_{\rm C}}{\eta^\star} - 1 + O\left(\frac{1}{\tau}\right)\,,
\label{eq:deta_largedP}
\end{equation}
and thus $\eta = \eta_{\rm C} + O(1/\tau)$. This means that, for large $\tau$, the efficiency depends on the parameter $a$ only via the term proportional to $1/\tau$, which becomes negligible close to $\delta_P\to-1$.
 
The second analytically tractable situation, which is more important for practical reasons, is the case of small $\delta_P$. Close to the maximum power the parameters $a$ and $\tau$ are also small. This means that, instead of performing the derivation for a small $\delta_P$, one can perform it for a small $a$. Data from the exact numerical optimization shown in Fig.~\ref{fig:universal} demonstrate that the absolute values of the optimal parameter $a$ are always either small (for moderately small $|\delta_P|$) or close to $-1$ (for $\delta_P \to -1$, when $\tau\gg 1$). This means that the optimization using the small $a$ approximation may be close to the exact solution even for relatively large $\delta_P$. This is because the effect of $a$ on the optimal efficiency is either well captured by the approximation (for moderate $\delta_P$) or negligible ($\delta_P \to -1$, when $\tau\gg 1$). Up to the second order in $a$, it follows from Eq.~(\ref{eq:tau_exact}) 
\begin{equation}
\tau = \pm\frac{\sqrt{-\delta_P}}{1 \mp \sqrt{-\delta_P}}\mp \frac{a^2}{2 A \sqrt{-\delta_P}}\,,
\label{eq:t_small_a}
\end{equation}
where the upper signs correspond to the upper sign in Eq.~(\ref{eq:tau_exact}) and thus lead to the maximum efficiency for the fixed power. The rest of the calculation can be performed without any other approximation. The final results are depicted in Fig.~\ref{fig:universal} by the gray lines, which in the panels (a) (maximum $\delta_\eta$ for a fixed power), (b) (maximum $\eta$ for a fixed power) and (c) (the corresponding optimal parameter $\tau$) overlap with the data obtained using exact numerical optimization. The only difference between the approximate analytical solution and the numerical results can be observed in panel (d), where we show the optimal values of the parameter $a$. Thus, as we have conjectured above, the results based on the approximate Eq.~(\ref{eq:t_small_a}) if no other approximations 
are made describe very well the exact optimized values of $\eta$ and $\delta_\eta$. Nevertheless, the formulas are quite involved and thus we will write in the rest of this section only the results up to the leading order in $\delta_P$.

Substituting $\tau$ with the upper signs from Eq.~(\ref{eq:t_small_a}) into Eq.~(\ref{eq:reldeta}) for $\delta_{\eta}$, taking the derivative with respect to $a$ and solving the resulting equation $d \delta_{\eta}/da=0$ for $a$, we obtain in the leading order in $\delta_P$
\begin{equation}
a = -\frac{1}{2}\frac{A\, \eta_{\rm C} }{1 + A - \eta_{\rm C}}\sqrt{-\delta_P}\le 0\,.
\label{eq:a_small_a}
\end{equation}
The resulting optimal parameter $a$ is thus negative ($\alpha<\alpha^\star$) in accord with the discussion at the end of Sec.~\ref{sec:Num_res}. Inserting $\tau$ from Eq.~(\ref{eq:t_small_a}) and $a$ from Eq.~(\ref{eq:a_small_a}) into the formula (\ref{eq:reldeta}) for $\delta_{\eta}$, we get up to the leading order in $\delta_P$
\begin{equation}
\delta_\eta = f(A,\eta_{\rm C}) \sqrt{-\delta_P}\,,
	\label{eq:deta_small_a}
\end{equation}
where
\begin{equation*}
f(A,\eta_{\rm C}) = \frac{1}{4} \left[\frac{(A+1) A}{A-\eta_{\rm C}+1}+\frac{4 (A+1) (A+2)}{-2 A+\eta_{\rm C}-2}+A+8\right]\,.
\label{eq:fAetaC}
\end{equation*}
The corresponding maximum efficiency $\eta = (\delta_\eta + 1)\eta^\star$ reads
\begin{equation}
\eta = \eta^\star(A,\eta_{\rm C})\left[1+f(A,\eta_{\rm C}) \sqrt{-\delta_P}\right]\,.
\label{eq:efficiency_small_a}
\end{equation}
Equations (\ref{eq:deta_small_a})--(\ref{eq:efficiency_small_a}) constitutes our first main result. The maximum relative gain in efficiency (\ref{eq:deta_small_a}) and the maximum efficiency itself (\ref{eq:efficiency_small_a}) are non-analytical functions of $\delta_P$ with a diverging slope at $\delta_P = 0$, which clearly points out that the gain in efficiency when working near maximum power is much larger then the loss of power, in accord with the findings of \cite{Holubec2015}. Both the diverging slope with $\delta_P\to 0$ and the scaling $\sqrt{-\delta_P}$ are direct consequences of the fact that the power has maximum at $\delta_P = 0$ and thus represent generic features of the maximum efficiency close to maximum power. 

In order to understand how these results arise, assume that both power, $P$, and the corresponding maximum efficiency, $\eta$, are parametrized by the parameter vector $\mathbf x$, in the present setting $\mathbf x = \{t_{\rm p},\alpha\}$, and that they are analytical functions of all these parameters. Taylor expansions of $P$ and $\eta$ around the point of maximum power ${\mathbf x}={\mathbf x}^\star$ [where $P' = \nabla P|_{{\mathbf x}={\mathbf x}^\star} = 0$ denotes the gradient and $P'' = \nabla^2 P|_{{\mathbf x}={\mathbf x}^\star}<0$ the negative definite Hessian matrix evaluated at the point of maximum power], $P = P^\star + ({\mathbf x}-{\mathbf x}^\star)^{\rm T} P''({\mathbf x}-{\mathbf x}^\star)/2$ and $\eta = \eta^\star + ({\mathbf x}-{\mathbf x}^\star)^{\rm T} \eta'$, lead to $\delta_\eta \propto \sqrt{-\delta_P}$. The scaling (\ref{eq:deta_small_a}) is thus universal whenever the used Taylor expansions of power and efficiency are valid. Indeed, the dependence (\ref{eq:deta_small_a}) has been already obtained for quantum thermoelectric devices \cite{Whitney2014,Whitney2015}, for a stochastic heat engine based on the underdamped particle diffusing in a parabolic potential \cite{Dechant2016} and also using linear response theory \cite{Ryabov2016}.  The next two terms in Eq.~(\ref{eq:deta_small_a}) are of the order $\delta_P$ and $(-\delta_P)^{3/2}$ and can be also accurately predicted if one departs from the approximate formula (\ref{eq:t_small_a}) for $\tau$.

\subsection{Maximum $\delta_\eta$ and $\eta$ as functions of the parameter $A$}
\label{Sec:Abeh}

The optimal relative gain in efficiency (\ref{eq:deta_small_a}) is an \emph{increasing} function of $A$ as can be proven by showing positivity of the derivative
\begin{equation}
\frac{\partial f(A,\eta_C)}{\partial A} = \eta_{\rm C}\frac{g(A,\eta_{\rm C})}{4(1 + A - \eta_{\rm C})^2 (-2 - 2 A + \eta_{\rm C})^2} A\,.
\label{eq:fder}
\end{equation}
The sign of this function is determined by the sign of the function $g(A,\eta_{\rm C}) = 8 (1 + A)^2 - 2 (1 + A) (7 + A) \eta_{\rm C} + 5 \eta_{\rm C}^2 + \eta_{\rm C}^3$.
The derivative of this expression with respect to $A$, $16(1+A)-4(1+A)\eta_{\rm C}-12\eta_{\rm C}$, is positive for all $\eta_{\rm C}$, $0< \eta_{\rm C} < 1$. The function $g(A,\eta_{\rm C})$ is thus an increasing function of $A$ and hence we can demonstrate that the positivity of $g(A,\eta_{\rm C})$ by showing that $g(0,\eta_{\rm C})>0$. To this end, we obtain $g(0,\eta_{\rm C}) = 8-14\eta_{\rm C} + 5\eta_{\rm C}^2+\eta_{\rm C}^3$. This expression decreases with $\eta_{\rm C}$ and thus the function $g(A,\eta_C)$ fulfills the inequality $g(A,\eta_C) > g(0,1) = 0$, which proves positivity of the derivative $\partial f(A,\eta_C)/\partial A$. Therefore, for small values of $\delta_P$, the maximum relative gain in efficiency for a given power increases with $A$. Furthermore, the same can be inspected from the full solution for the optimal $\delta_\eta$, using the exact numerical optimization and also using the analytical results for $\delta_P\to -1$ (\ref{eq:deta_largedP}). This means that the limit $A\to 0$ of $\delta_\eta$ yields the lower bound on the relative gain in efficiency for arbitrary $\delta_P$. The upper bound on $\delta_\eta$ is then obtained in the limit $A\to \infty$. 

Similar argumentation can be used also for the optimal efficiency at a given power. For small values of $\delta_P$ the optimal $\eta$ is a monotonously \emph{decreasing} function of $A$ as can be shown using the equation (\ref{eq:efficiency_small_a}). According to this equation the derivative of the maximum efficiency with respect to $A$ is given by 
\begin{equation}
\frac{\partial \eta}{\partial A} = \frac{\partial \eta^\star}{\partial A} + \left(\frac{\partial \eta^\star}{\partial A} f + \eta^\star \frac{\partial f}{\partial A}\right) \sqrt{-\delta_P}\,.
\label{eq:etader}
\end{equation}
 As can be inspected directly from its definition (\ref{eq:eta_opt}), $\eta^\star$ decreases with $A$, i.e. $\partial\eta^\star/\partial A < 0$. This means that $\partial\eta/\partial A < 0$ and the maximum efficiency decreases with $A$ for small values of $\delta_P$. Furthermore, the same behavior, but now for arbitrary $\delta_P$, is obtained using the full solution for the optimal efficiency and also using the exact numerical optimization. Finally, for $\delta_P\to-1$ the maximum efficiency equals to $\eta_{\rm C}$ for any $A$. The lower bound for the optimal efficiency is thus obtained for $A\to \infty$ and corresponds to the upper bound for the optimal $\delta_P$. Similarly, the upper bound for the optimal $\eta$ is obtained for $A\to 0$ and corresponds to the lower bound for the optimal $\delta_P$. 

Physically, this behavior can be understood if one realizes how the quantity $A_{\rm c}$ contributes to the total entropy production $\Delta S_{\rm tot}$.
At the end of Sec.~\ref{sec:Num_res} we argued that, by decreasing $\alpha$, larger part of the total dissipated work is delivered to the hot bath, where it produces smaller amount of entropy than it would produce in the cold reservoir. For a fixed power, the parameters $A_{\rm c}$ and $A_{\rm h}$ are no longer independent since they satisfy Eq.~(\ref{eq:power_Wirr}). By changing these parameters one redistributes the total amount of dissipated work between the two reservoirs in the same way as by changing the parameter $\alpha$. If the parameter $A_{\rm c}$ is small, larger amount of work is dissipated in the hot bath and, similarly, for a large $A_{\rm c}$ more work is dissipated in the cold bath. This means that the efficiency decreases (entropy production increases) with increasing $A = \sqrt{A_{\rm c}/A_{\rm h}}$ and vice versa. 

Does this also imply that larger $A$ lead to larger gain in efficiency $\delta_\eta = \eta/\eta^\star-1$? As we have argued above, both the EMP $\eta^\star$ and the maximum efficiency at a given power $\eta$ are decreasing functions of $A$. The fact that $\delta_\eta$ is an increasing function of $A$ means that the decrease of $\eta$ with $A$ must be slower than the decrease of $\eta^\star$. The EMP $\eta^\star$ is completely determined by the condition that the corresponding power is maximal (parameters $a$ and $\tau$ are fixed) and thus it has no freedom to be further optimized when the parameter $A$ changes. On the other hand, the maximum efficiency $\eta$ at a given power possesses such freedom and thus one may expect, that it will decay with increasing $A$ slower than $\eta^\star$. Our results for behavior of optimal $\eta$ and $\delta_\eta$ with $A$ verify this conjecture (see Fig.~\ref{fig:dP_eta}). Now, let us focus on deriving the bounds for maximum gain in efficiency for a given power and for the maximum efficiency for a given power.

\section{Bounds on maximum gain in efficiency}
As we have discussed in Sec.~\ref{Sec:Abeh}, the upper bound on $\delta_{\eta}$ follows by taking the limit $A\to\infty$ in Eqs.~(\ref{eq:reldP})--(\ref{eq:reldeta}). The result is
\begin{eqnarray}
\lim_{A\to \infty}\delta_{P} &=& -\left(\frac{\tau}{1+\tau}\right)^2\,,
\label{eq:dP_Binf}\\
\lim_{A\to \infty}\delta_{\eta} &=& \frac{\tau}{1+\tau} = \sqrt{-\delta_P}\,.
\label{eq:deta_Binf}
\end{eqnarray}
The lower bound follows by taking the other total asymmetric limit $A\to 0$. Then $\alpha^\star = 1$ and thus $a \in [-1,0]$. From Eq.~(\ref{eq:tau_exact}) we get
\begin{equation}
\tau = \frac{-\delta_P}{1+\delta_P} \pm \frac{\sqrt{a-\delta_P}}{\sqrt{1+a}(1+\delta_P)}\,,
\label{eq:t_A0}
\end{equation}
where, for $a \in [-1,0]$, $a - \delta_P > 0$ as can be shown directly from Eq.~(\ref{eq:reldP}). Positive relative change in efficiency 
\begin{equation}
\delta_\eta = \frac{2(1-\eta_{\rm C})\left[a-\delta_P+\sqrt{(1 + a) (a - \delta_P)}\right]}{2 \left[a + 1 + \sqrt{(1 + a) (a - \delta_P)}\right] - (1 + \delta_P) \eta_{\rm C}}
\label{eq:deltaeta_A0_pre}
\end{equation}
is obtained for the plus sign before the square root in Eq.~(\ref{eq:t_A0}). From $a \in [-1,0]$ and $a - \delta_P > 0$ it follows that $\partial \delta_\eta / \partial a > 0$
and thus $\delta_\eta$ monotonously increases with $a$. This means that the maximum
\begin{equation}
\delta_\eta =\frac{2(1-\eta_{\rm C})\sqrt{-\delta_P}}{2-(1-\sqrt{-\delta_P})\eta_{\rm C}}
\label{eq:deltaeta_A0}
\end{equation}
is obtained for maximum possible value of $a$, $a=0$.

We have thus found that the maximum gain in efficiency at a given power obeys the inequalities
\begin{equation}
\frac{2(1-\eta_{\rm C})\sqrt{-\delta_P}}{2-(1-\sqrt{-\delta_P})\eta_{\rm C}} \le \delta_\eta \le \sqrt{-\delta_P}\,.
\label{eq:deta_bounds2}
\end{equation}
As we have discussed at the end of Sec.~\ref{Sec:Abeh}, the upper bound (\ref{eq:deta_bounds2}) corresponds to the lower bound on maximum efficiency at a given power, $\eta = (1+\delta_{\eta})\eta^\star$, and, similarly, the lower bound (\ref{eq:deta_bounds2}) yields the upper bound on $\eta$. For $A\to\infty$, we have $\eta^\star\to \eta_{\rm C}/2$ and for $A\to 0$, $\eta^\star\to \eta_{\rm C}/(2-\eta_{\rm C})$. The bounds on efficiency thus read
\begin{equation}
\frac{\eta_{\rm C}}{2}\left(1+\sqrt{-\delta_P}\right) \le \eta  \le \eta_{\rm C}\frac{1 + \sqrt{-\delta_P}}{2 - (1-\sqrt{-\delta_P})\eta_{\rm C}}\,.
\label{eq:eta_bounds}
\end{equation}
The bounds (\ref{eq:deta_bounds2})--(\ref{eq:eta_bounds}) are our second main result. They represent direct generalization of the bounds on EMP derived for $\delta_P = 0$ by Esposito et al. \cite{Esposito2010b}. Note that for small temperature differences, i.e. up to the leading order in $\eta_{\rm C}$, both the lower and the upper bound on the maximum efficiency equal and thus the maximum efficiency as a function of $\delta_P$ is independent of the parameter $A$, which contains details about the system dynamics. It is given by
\begin{equation}
\eta = \frac{\eta_{\rm C}}{2}\left(1+\sqrt{-\delta_P}\right)\,.
\label{eq:deta_small_eC}
\end{equation}
The same formula for maximum efficiency has been recently obtained using linear response theory in the strong coupling limit \cite{Ryabov2016}.

In Fig.~{\ref{fig:universal}}(a) we show the bounds (\ref{eq:deta_bounds2}) and in Fig.~{\ref{fig:universal}}(b) we show the corresponding bounds on the maximum efficiency (\ref{eq:eta_bounds}). From the figure, one can inspect that the maximum efficiency interpolates between the EMP $\eta^\star$ (for $\delta_P=0$) and Carnot efficiency $\eta_{\rm C}$ (for $\delta_P=-1$), which is, in accord with the bounds (\ref{eq:eta_bounds}), reached irrespectively of the parameter $A$. Similar behavior of maximum efficiency was encountered for the underdamped particle diffusing in a parabolic potential \cite{Dechant2016}.

\section{Conclusions and outlooks}
It is well known that real-world heat engines should not work at maximum power, but rather in a regime with slightly smaller power, but with considerably larger efficiency. For low-dissipation heat engines, we have introduced lower and upper bounds on the maximum efficiency at a given power (\ref{eq:deta_bounds2}) and the corresponding bounds on the maximum efficiency (\ref{eq:eta_bounds}). We have also calculated maximum relative gain in efficiency for arbitrary fixed power. Close to maximum power, this gain scales as a square root from the relative loss of power $\delta_P$ (\ref{eq:deta_small_a}). This scaling is a direct consequence of the fact that power has maximum at $\delta_P = 0$ and thus it is universal for a broad class of systems. Indeed, the same scaling of maximum efficiency with the relative loss of power has been found recently for several models \cite{Whitney2014,Whitney2015,Dechant2016,Ryabov2016}.
Our results thus support the general statement about actual heat engines with quantitative arguments and reveal more practical limits on efficiency than the reversible one.

It would be interesting to investigate maximum gain in efficiency for a fixed power also for other models, such as endoreversible heat engines, or systems described by general Markov dynamics, i.e. by a Master equation, to see whether the behavior would be qualitatively the same as that obtained here and in the studies \cite{Whitney2014,Whitney2015,Dechant2016,Ryabov2016}. Furthermore, one can ask if the functional form of the prefactor $f$ in the formula for the gain in efficiency $\delta_\eta = f \sqrt{-\delta_p} + O(\delta_p)$ is controlled by similar symmetries of the underlying dynamics as the EMP \cite{Esposito2009,Izumida2014,Sheng2015,Cleuren2015}. It would be also immensely interesting to find a heat engine where the square root scaling of the maximum gain in efficiency close to maximum power would not be valid.


%


\end{document}